\documentstyle[12pt]{article}
\newcommand{\be}{\begin{equation}}
\newcommand{\en}{\end{equation}}
\newcommand{\ba}{\begin{eqnarray}}
\newcommand{\ea}{\end{eqnarray}}
\newcommand{\half}{{\scriptstyle{\frac{1}{2}}}}
\def\eqn#1{~(\ref{#1})}

\def\half{{\scriptstyle \frac{1}{2}}}
\begin{document}
\begin{center}
Contribution to the Workshop on Quantum Infrared Physics \\
     American University, Paris, 6-10 June, 1994\\

\bigskip
{\bf HORIZON CONDITION HOLDS POINTWISE\\
ON FINITE LATTICE WITH FREE BOUNDARY CONDITIONS}
\bigskip

MARTIN SCHADEN\footnote{Research supported by Deutsche
Forschungsgemeinschaft under grant no. Scha/1-1} and DANIEL
ZWANZIGER\footnote{Research supported by the National Science
Foundation under grant no. PHY93-18781}\\
Physics Department, New York University\\
New York, N. Y. 10003, U.S.A.\\
\bigskip

\end{center}

\centerline{ABSTRACT}
\begin{list}%
{}{\setlength{\leftmargin}{3pc}\setlength{\rightmargin}{\leftmargin}}
\small\setlength{\baselineskip}{12pt}
\item It is proven that the "horizon condition", which was found to
characterize the fundamental modular region in continuum theory and
the thermodynamic limit of gauge theory on a periodic lattice, holds
for every (transverse) configuration on a finite lattice with free
boundary conditions.
\end{list}

\section{\normalsize\bf Introduction}
	In his seminal work, Gribov$^{1}$ pointed out that the
non-triviality of the fundamental modular region in Coulomb or Landau
gauge has dynamical consequences.  [The fundamental modular region is
defined explicitly in Eq.\eqn{4} below.]  Recently$^{2}$, the
fundamental modular region was studied in the infinite-volume or
thermodynamic limit of a periodic lattice.  It was found that the
fundamental modular region is characterized by a "horizon condition"
in the sense that the Euclidean probability gets concentrated where
the horizon function $H(U)$, a bulk quantity of order $V$, defined in
Eq.\eqn{39} below, vanishes.  More precisly, it was shown
that  at large volumes the vacuum expectation value of $H(U)$ is of
order unity, $\langle H(U)\rangle=O(1)$, (instead of $V$) that its
variance is of order $V$ , $\langle H^2(U)\rangle=O(V)$, instead of $V^2$.
It was also found$^{2}$ that the horizon condition may be implemented by a
Boltzmann factor $\exp[-\alpha H(U)]$, where $\alpha$ is a
thermodynamic parameter determined by the constraint $\langle
H(U)\rangle=0$.   Recently a
calculation of glueball masses has been carried out$^{3}$ in which
the mass scale is set by $\alpha$.  Previous calculations of glueball
masses from the
fundamental modular region of gauge theory have been done using a mode
expansion by
Cutkosky and co-workers$^{4}$ and by Van Baal  and co-workers$^{5}$.

	The argument whereby the horizon condition is established in
the {\it infinite-volume}  limit of the {\it periodic} lattice, relies on two
technical hypotheses$^{2}$. In the present contribution, we shall prove
that, remarkably, the horizon condition holds {\it point-wise} for every
transverse configuration on the {\it finite} lattice with {\it free boundary
conditions}, $H(U)=0$.

\section{\normalsize\bf Lattice with Free Boundary Conditions in Minimal Landau
Gauge}

We consider a finite hypercubic lattice without
periodicity conditions, in $D$ Euclidean space-time dimensions.
Along each principal axis of the lattice there are $L$ links and $L+1$
sites. 	Lattice configurations $U$ are described by link
variables $U_{xy}=U_{yx}^\dagger$, and local gauge transformations $g$ by
the site variables $g_x$, where $U_{xy}$ and $g_x$ are both elements
of $SU(n)$.
The gauge-transform $U^g$ of the configuration $U$ by a local gauge
transformation $g$, is given explicitly by
\be\label{1}
(U^g)_{xy}=g^\dagger_x U_{xy} g_y\,.
\en

For the purposes of analytic calculations, it is convenient to
minimize gauge fluctuations by choosing a gauge which makes the link
variables $U_{xy}$, on each link $(xy)$ of the lattice, as close to unity as
possible, in an equitable way over the whole lattice.  For this
purpose, we take as a measure of the deviation of the link variables
from unity, the quantity
\be\label{2}
I(U) = \sum_{(xy)} n^{-1} {\rm Re\ tr}(1 -  U_{xy}) \,,
\en
where the sum extends over all links $(xy)$ of the lattice.  It is
positive, $I(U)\geq 0$, and vanishes, $I(U)=0$, if and only
if $U_{xy}=1$ on every link $(xy)$.  [The continuum analog of this
expression is the Hilbert norm of the connection $A$, $I[A] = \int
{\rm d}^D x |A(x)|^2$.]  The restriction of this quantity to the gauge
orbit through an arbitrary configuration $U$ is given by the Morse function
\be\label{3}
F_U(g)\equiv I(U^g)\,,
\en
regarded as a function on the local gauge group $G$, for fixed $U$.  Gauge
fluctuations are minimized by the gauge-fixing which consists in
choosing as the representative, on each gauge orbit, that
configuration $U$ which yields the absolute minimum of $F_U(g)$ at
$g_x=1$.  The set
$\Lambda$ of all configurations $U$ such that this function is an absolute
minimum on each gauge orbit at $g_x=1$, constitutes the
fundamental modular region,
\be\label{4}
\Lambda\equiv \{U:  F_U(1)\leq F_U(g)\ {\rm for\ all}\ g\}\,.
\en
Degenerate absolute minima occur only on the boundary of the
fundamental modular region, and are identified topologically.

At any local or absolute minimum, the minimizing function is
stationary,  which gives the local gauge condition. The second
variation is non-negative.  To obtain the explicit form of these
quantities, we write the local gauge transformation
$g_x=\exp (\omega_x)$, where $\omega_x=\omega_x^a t^a$.  Here the $t^a$ form an
anti-hermitian basis of the defining representation of the Lie algebra
of $SU(n)$, $[t^a, t^b]=f^{abc} t^c$, normalized to
${\rm tr}(t^a t^b)=-\half\delta^{ab}$.  We have, to second order in $\omega$,
\ba\label{5}
F_U(g) &=& F_U(1) - (2n)^{-1}\sum_{(xy)}[{\rm tr}\{ [U_{xy} -
U_{xy}^\dagger] (\omega_y - \omega_x + \half[\omega_x, \omega_y])
\}\nonumber\\
& &\mbox{}{\hskip 10em} +\half {\rm tr}\{[U_{xy} + U_{xy}^\dagger ] (\omega_y -
\omega_x)^2 \}
]\nonumber\\
& &\mbox{}\nonumber\\
F_U(g) &=& F_U(1) + (2n)^{-1}\sum_{(xy)}[A_{xy}^c (\omega_y^c - \omega_x^c +
\half f^{abc} w^a w^b)\nonumber\\
& &\mbox{}{\hskip 10em} +\half (\omega_y^a - \omega_x^a) G_{xy}^{ac}
(\omega_y^c - \omega_x^c) ]\,.
\ea
Here we have introduced the real link variables
\be\label{6}
A_{xy}^c\equiv - {\rm tr}[ (U_{xy} - U_{xy}^\dagger) t^c ]\,,
\en
which approach the classical connection in the continuum limit, and
\be\label{7}
G_{xy}^{ab}\equiv - \half{\rm tr}\{ (t^a t^b + t^b t^a ) [ U_{xy} +
U_{xy}^\dagger ] \}\,.
\en
We introduce some elementary geometry of this lattice that
will be useful.  Let $\phi_x\equiv \phi(x)$ be a site variable, and
let $(xy)$ denote a link of a lattice. Corresponding link
variables are defined by
\be\label{8}
(\nabla \phi)_{xy}\equiv \phi_x - \phi_y
\en
\be\label{9}
(a\phi)_{xy}\equiv\half (\phi_x + \phi_y)\,.
\en
These are defined for all links, including those for which $x$ or $y$ may
lie on a face, edge or vertex.  (For $\nabla\phi$ we could alternatively use
the Cartan notation d$\phi$).  We may also uniquely designate links by
$(xy)=(x,\mu)$, for $y=x+e_\mu$, where $e_\mu$ points in
the positive $\mu$-direction, and we shall also use, as convenient, the
notation for link variables
\be\label{10}
\nabla_\mu\phi(x)\equiv (\nabla\phi)_{yx}
\en
\be\label{11}
a_\mu\phi(x)\equiv (a\phi)_{xy}
\en
(These expressions are \underline{undefined} for sites $x$ where
$y=x+e_\mu$ is not a site of the lattice.)  Given two link
variables $V_{xy}$ and $W_{xy}$, we may form the link variable
$P_{xy}=V_{xy}W_{xy}$, etc.

	We also introduce the lattice divergence of a link variable
$A_\mu(x)$, which is a site variable $(\nabla\cdot A)_x$, defined by the dual
\be\label{12}
(\nabla\cdot A, \phi)\equiv - (A, \nabla\phi) = - \sum_{(xy)} A_{xy}
(\nabla \phi)_{xy} = \sum_x (\nabla\cdot A)_x \phi_x\,,
\en
where the sums extend over all links and sites of the lattice.
Similarly the site variable $(a\cdot A)_x$ is defined by
\be\label{13}
(a\cdot A, \phi)\equiv (A, a\phi) = \sum_{(xy)} A_{xy} (a\phi)_{xy} = \sum_x
(a\cdot A)_x \phi_x\,.
\en
The lattice divergence is defined for all sites $x$, and represents the
sum of all link variables leaving the site $x$.  (It is {\it not} formed
simply of differences of link variables when $x$ is a boundary point of
the lattice.)

With the help of these definitions, we rewrite Eq.\eqn{5} in the form
\be\label{14}
F_U(g) = F_U(1) + (2n)^{-1} [ - (A, \nabla \omega) +  (\nabla \omega,
D(U)\omega)
]\,.
\en
Here $[D(U)\omega]_{yx}$ is a well-defined link variable which we call the
lattice gauge-covariant derivative of the site variable $\omega_x$,
\be\label{15}
D_\mu^{ac} \omega^c(x)=
G_\mu^{ac}(x)\nabla_\mu\omega^c(x)+f^{abc}A_\mu^b(x)a_\mu
\omega^c(x)\,.
\en
It is the infinitesimal change in $A_\mu(x)$ under an infinitesimal gauge
transformation $\omega_x$.  We also define the lattice Faddeev-Popov
matrix $M(U)$  by
\be\label{16}
(\omega, M(U)\phi) = (\nabla\omega, D(U)\phi)\,.
\en

At a minimum, the minimizing function $F_U(g)$ is stationary, namely
$(A,\nabla\omega)=-(\nabla\cdot A,\omega)=0$.  This holds
for all $\omega$, so the gauge condition is expressed by the
transversality of A,
\be\label{17}
\nabla\cdot A=0\,.
\en
Because of transversality, the gauge just defined falls into the class
of lattice Landau gauges, and we call it the "minimal Landau gauge".
At a stationary point of the minimizing function, the matrix of second
derivatives is symmetric, so when $A$ is transverse, the Faddeev-Popov
matrix is symmetric,
\be\label{18}
M \equiv -\nabla\cdot D(U) =  - D(U)\cdot\nabla\qquad ({\rm for\ }
\nabla\cdot A=0)\,.
\en
This matrix is non-negative at a minimum, and the two conditions
together define the Gribov region $\Omega$,
\be\label{19}
\Omega\equiv \{U:  \nabla\cdot A(U) = 0\ {\rm and}\ M(U)\geq 0 \}\,.
\en

Because the set $\Lambda$ of absolute minima is contained in the set of
relative mimima, we have the inclusion
\be\label{20}
\Lambda\subseteq\Omega\,,
\en
where $\Lambda$ is the fundamental modular region.

\section{\normalsize\bf Vanishing of the Horizon Function on the Finite Lattice
with Free
Boundary Conditions}

Because there are $L+1$ sites, but only $L$ links on each row of
the lattice with free boundary conditions, the transversality
condition is more restrictive for free boundary conditions than on a
periodic lattice, so various additional constraints hold, as we shall
now show.

Choose a particular direction, $\mu=0$, on the hypercubic
lattice, and call the corresponding (Euclidean) coordinate $t$ and
the other coordinates ${\bf x}$.  Consider the sum over the hyperplane
labelled by $t$, for a transverse configuration, $\nabla\cdot A =0$,
\be\label{21}
\sum_{\bf x} (\nabla\cdot A)(t,{\bf x}) = 0\,.
\en

Contributions to this sum from links that lie within the hyperplane
vanish, because all such links are connected to 2 sites within the
hyperplane and these 2 contributions cancel.  There remain the
contributions from perpendicular links.  For all $t$ that label interior
hyperplanes $1\leq t< L-1$, they give
\be
Q_0(t) - Q_0(t-1) = 0\,,\nonumber
\en
whereas on the boundary hyperplanes they give
\be
Q_0(L-1) = Q_0(0)= 0\,.\nonumber
\en
where
\be\label{22}
Q_0(t)\equiv\sum_{\bf x} A_0(t,{\bf x})\,.
\en
We thus have proved the \underline{lemma}:\hfill\\
Let $A_\mu(x)$ be a transverse configuration, $\nabla\cdot A(x)=0$, on a
lattice with free boundary conditions.  Then  the "charges" vanish,
\be\label{23}
Q_0(t)\equiv\sum_{\bf x} A_0(t, {\bf x}) = 0\qquad (t = 0,1,\dots L-1)\,.
\en
[For a lattice with periodic boundary conditions, only the weaker
condition $Q_0(t)={\rm const.}$ holds.]  If we sum the last equation
over $t$, for generic $\mu$, we obtain:\hfill\\
\underline{Corollary}.  Let $A_\mu(x)$ be a transverse configuration
$\nabla\cdot A(x)=0$ on a lattice with free boundary conditions.
Then the zero-momentum component of A vanishes,
\be\label{24}
\sum_x A_\mu(x) = 0\,.
\en
Note that for the periodic lattice, transversality imposes no
condition whatsoever on the constant component of $A_\mu(x)$.  On the other
hand, for the periodic lattice, the core of the fundamental modular
region was found$^{2}$ to satisfy Eq.\eqn{24}.

We shall now prove that the horizon condition is satisfied
point-wise for a finite lattice with free boundary conditions.
Consider the integral
\ba\label{25}
I &\equiv&\int {\rm d}\phi {\rm d}\phi^* \exp[ ( \phi^*_i, M(U)\phi_i)
]\,,\nonumber\\
&=& \int {\rm d}\phi {\rm d}\phi^* \exp[ - (\nabla_\lambda\phi^{*a}_i,
D_\lambda^{ac}(U)\phi_i^c) ]\,,
\ea
where the $\phi_i^a(x)$ variables are integrated over the real axis,
and the $\phi^{*a}_i(x)$ are independent variables that are integrated over the
imaginary axis. The lattice gauge-covariant derivative $D_\lambda^{ac}(U)$ acts
on the upper index, of $\phi_i^c(x)$, and $i=1,\dots f$ is a dummy
index.  For transverse $A=A(U)$, as we assume, the
symmetric matrix $M(U)\equiv-\nabla\cdot D(U)=-D(U)\cdot\nabla$ has a
null-space, $H_0$,  consisting of
constant functions, so the integral\eqn{25} would diverge if $\phi(x)$
and $\phi^*(x)$
were independent integration variables for every $x$.  The
integral is made finite by the constraint at a fixed point $y$ of the
lattice
\be\label{26}
\phi_i^c(y) = \phi_i^{*c}(y) = 0\,,
\en
and the definition
\be\label{27}
{\rm d}\phi {\rm d}\phi^*\equiv \prod_{x\not=y, i, a} {\rm
d}\phi_i^a(x) {\rm d}\phi^{*a}_i(x)\,.
\en
With this specification, the integral has the value
\be\label{28}
I=\int {\rm d}\phi\prod_{i=1}^f\delta(M\phi_i) = {\det}^{-f}M_y\,,
\en
where $M_y$ is the matrix obtained from $M$ by deleting the rows and
columns that bear the label $y$.

Because $M(U)$ has a null space, $H_0$, consisting of constant
functions $\nabla\omega=0$, an alternative expression for $I$ is
\be\label{29}
I= {\rm d}\phi {\rm d}\phi^* \exp[ (\phi^*_{i\perp},
M(U)\phi_{i\perp}) ]\,,
\en
where $\phi_\perp$ and $\phi^*_\perp$ are the projections of $\phi$
onto the orthogonal subspace $H_\perp$,
\be\label{30}
\phi_\perp (x)= \phi(x)- S^{-1}\sum_x\phi(x)\,,
\en
and $S$ is the total number of sites in the lattice.  The variables $\phi(x)$
and $\phi^*(x)$ for $x\not=y$ constitute a complete set of (linear)
coordinates on $H_\perp$, and the change of basis\eqn{30} is configuration
independent.  We conclude that
\be\label{31}
I = {\det}^{-f}M_y = {\det}^{-f}M_\perp\,,
\en
which shows that $I$ is independent of $y$.  (The equation
${\det} M_y={\det} M_\perp$,holds for any symmetric matrix $M$ with the
property
that the sum of each row and each column vanish.)

Now let the dummy index $i$ represent the pair
$i=(\mu,a)$, so $\phi_{\mu,a}^c(x)$ is a site variable with a preferred
direction $\mu$, and make the shift on $\phi$ and $\phi^*$ given by
\ba\label{32}
\phi_{\mu,a}^c(x) &=& \phi_{\mu,a}^{\prime c}(x) + x_\mu
\delta^c_a\nonumber\\
\phi_{\mu,a}^{*c}(x) &=& \phi_{\mu,a}^{*\prime c}(x) + x_\mu \delta^c_a
\ea
which are the lattice analogs of shifts previously introduced in
continuum theory$^{6}$.  This shift makes no sense on a (finite) periodic
lattice, but it is well defined on a finite lattice with free boundary
conditions. (Note that the constraint at the lattice point $y$ for the
shifted fields differs from\eqn{26}, unless it is imposed at the
``origin'' $y=0$.) For $\nabla\cdot A=0$, one finds after a simple
calculation
\ba\label{33}
I&\equiv&\int {\rm d}\phi {\rm d}\phi^* \exp[
(\nabla_\lambda\phi^{*b}_{\mu,a}, D_\lambda^{bc}(U)\phi_{\mu,a}^c) +
(D_\lambda^{ac}(U)\phi_{\lambda,a}^c)\nonumber\\
& &\mbox{}{\hskip 10em}+(D_\lambda^{ac}(U)\phi^{*c}_{\lambda,a}) +
\sum_{x,\mu} G_\mu^{aa}(x) ]\,.
\ea

It is convenient to introduce a field variable $B_{\lambda,a}^c(x)$,
by duality
\be\label{34}
( B_{\lambda,a}^c, \phi_{\lambda,a}^c)\equiv -
(D_\lambda^{ac}(U)\phi_{\lambda,a}^c)\, .
\en
Like $\phi_{\lambda,a}^c$, the new variable $B_{\lambda,a}^c$ is a
site variable with a preferred direction $\mu$.  We have
\be\label{35}
B_{\lambda,a}^c(x) = (\nabla\cdot G)_\lambda^{ac}(x) - f^{abc} (a\cdot
A)_\lambda^b(x)\,,
\en
where $(\nabla\cdot G)_\lambda^{ac}(x)$  represents the contribution
to the lattice divergence $\nabla\cdot G^{ac}(x)$ associated with the
$\lambda$ axis, and similarly
for $(a\cdot A)_\lambda^b(x)$. With this definition,
\ba\label{36}
I&\equiv&\int {\rm d}\phi {\rm d}\phi^* \exp[
(\nabla_\lambda\phi^{*b}_{\mu,a}, D_\lambda^{bc}(U)\phi_{\mu,a}^c) -
(B_{\lambda,a}^{c},\phi_{\lambda,a}^c)\nonumber\\
& &\mbox{}{\hskip 10em}-(\phi^{*c}_{\lambda,a},B_{\lambda,a}^c) +
\sum_{x,\mu} G_\mu^{aa}(x) ]\,.
\ea

We may now effect the $\phi$ and $\phi^*$ integrations by making the shifts
\ba\label{37}
\phi_{\lambda,a}^b &=& \phi_{\lambda,a}^{\prime b} +
(M^{-1})^{bc}B_{\lambda,a}^c \nonumber\\
\phi_{\lambda,a}^{*b} &=& \phi_{\lambda,a}^{*\prime b} + (M^{-1})^{bc}
B_{\lambda,a}^c\,.
\ea
This expression is well-defined because $B_{\lambda,a}^c$ is
orthogonal to the null-space of $M$,
\be\label{38}
\sum_x B_{\lambda,a}^c(x) = 0\,,
\en
and we may choose coordinates adapted to these subspaces.  To see
this, observe that $\sum_x(\nabla\cdot G)_\lambda^{ac}(x)$ vanishes
because each link
is connected to two sites and gives opposite contributions from each.
Moreover $\sum_x(a\cdot A)_\lambda^b(x) = \sum_x A_\lambda^b(x)$
vanishes by the preceding corollary.  This gives $I = I \exp[-H(U)]$,
where $H(U)$ is given by Eq.\eqn{39}.  We have proven:\hfill\\
\underline{Theorem}.  Let $U$ be a transverse configuration
$\nabla\cdot A(U)=0$,
and let the horizon function $H(U)$ on a finite lattice with free
boundary conditions be defined by
\be\label{39}
H(U) = (B_{\lambda,a}^b, (M^{-1})^{bc} B_{\lambda,a}^c) - \sum_{x,\mu}
G_\mu^{aa}(x)\,,
\en
where $B$ is defined in\eqn{35} and $G_\mu^{ab}(x)$ in\eqn{7}.  Then
$H(U)$ vanishes
\be\label{40}
H(U) = 0\,.
\en

\noindent\underline{Remark}.  On a periodic lattice, the horizon
condition does not hold point-wise.  For example, the vacuum
configuration $U_\mu(x)=1$ on
a {\it periodic} lattice gives $B_{\lambda,a}^c(x)=0$, so $H(U)=-(n^2-1)DV$.
For the vacuum configuration on the finite lattice with {\it free} boundary
conditions,  $B_{\lambda,a}^c(x)$ is entirely supported on the boundary.  Thus
boundary contributions remain important at arbitrary large volume.
Long range boundary effects have also been found by Patrascioiu and
Seiler$^{7}$.

\noindent\underline{Discussion}.  The result\eqn{40} is quite
remarkable and unexpected.
Recall that as the {\it volume of the periodic lattice approaches infinity},
the probability gets concentrated where the horizon condition holds$^{2}$
$H(U)=0$, and where $\sum_x A(x) = 0$.  Here we have just
proven that these conditions are satisfied for all transverse
configurations $U$ on the finite lattice with free boundary
conditions.

\section{\normalsize\bf References}
\newcounter{ref}
\begin{list}%
{[\arabic{ref}]}{\usecounter{ref}\setlength{\leftmargin}{\parindent}}

\item 	V. N. Gribov, Nucl. Phys. B139 (1978) 1.
\item  	D. Zwanziger, Nucl. Phys. B412 (1994) 657.
\item 	M. Schaden and D. Zwanziger, {\it Glueball Masses from
the Gribov Horizon: Basic Equations and Numerical Estimates}, NYU
preprint ThPhSZ94-1.
\item  	R. E. Cutkosky, J. Math. Phys. 25 (1984) 939; R. E. Cutkosky
and K. Wang, Phys. Rev. D37 (1988) 3024; R. E. Cutkosky, Czech J.
Phys. 40 (1990) 252.
\item 	P. van Baal and N. D. Hari Dass, Nucl. Phys B385 (1992) 185;
J. Koller and P. van Baal, Nucl. Phys. B302 (1991) 1; P. van Baal,
Acta Physica Pol. B20 (1989) 295; P. Van Baal and B. van den Heuvel,
{\it Zooming in on the SU(2) Fundamental Domain}, (preprint) University of
Leiden, INLO-PUB-12/93.
\item 	N. Maggiore and M. Schaden, {\it Landau Gauge within the Gribov
Horizon}, NYU preprint, October 1993, Phys. Rev. D (to be published).
\item  A. Patrascioiu and E. Seiler, {\it Super-Instantons in
Gauge Theories and Troubles with Perturbation Theory}, hep-lat
9402003, MPI-PhT/94-07, AZPH-TH/94-03, Phys. Rev. Lett (to be published).

\end{list}

\end{document}